\documentclass{article}
\usepackage{graphics,graphicx,epsfig,amsfonts,amsmath}
\usepackage{subfigure}
\begin{document}
\newcommand{\N}{\mathbb{N}}
\newcommand{\R}{\mathbb{R}}
\newcommand{\Q}{\mathbb{Q}}
\title{A non-linear mathematical model of cell turnover, differentiation and tumorigenesis in the intestinal crypt.}

\author{Alberto d'Onofrio $^{1}$ and Ian P. M. Tomlinson $^{2}$ \footnote{Current address: Nuffield Department of Medicine, Wellcome Trust Centre for Human Genetics, Oxford University, Oxford UK}
} 

\maketitle
%

{\center $^1$ Division of Epidemiology and Biostatistics,
 European Institute of Oncology, Via Ripamonti 435, Milano, Italy,
 I-20141\\EMAIL:
alberto.donofrio@ieo.eu\\ \center $^2$
Molecular and Population Genetics Laboratory, Cancer Research UK,
44, Lincoln's Inn Fields, London WC2A 3PX. UK  }

\begin{abstract}
We present a development of a model of the relationship between
cells in three compartments of the intestinal crypt: stem cells,
semi-differentiated cells and fully differentiated cells. Stem and
semi-differentiated cells may divide to self-renew, undergo
programmed death or �progress� to semi-differentiated and fully
differentiated cells respectively. The probabilities of each of
these events provide the most important parameters of the model.
Fully differentiated cells do not divide, but a proportion
undergoes programmed death in each generation. Our previous models
showed that failure of programmed death - for example, in
tumorigenesis - could lead either to exponential growth in cell
numbers or to growth to some plateau. Our new models incorporate
plausible fluctuation in the parameters of the model and introduce
non-linearity by assuming that the parameters depend on the
numbers of cells in each state of differentiation. We present
detailed analysis of the equilibrium conditions for various forms
of these models and, where appropriate, simulate the changes in
cell numbers. We find that the model is characterized by
bifurcation between increase in cell numbers to stable equilibrium
or 'explosive' exponential growth; in a restricted number of
cases, there may be multiple stable equilibria. Fluctuation in
cell numbers undergoing programmed death, for example caused by
tissue damage, generally makes exponential growth more likely, as
long as the size of the fluctuation exceeds a certain critical
value for a sufficiently long period of time. In most cases, once
exponential growth has started, this process is irreversible. In
some circumstances, exponential growth is preceded by a long
plateau phase, of variable duration, mimicking equilibrium: thus
apparently self-limiting lesions may not be so in practice and the
duration of growth of a tumor may be impossible to predict on the
basis of its size.
\end{abstract}

\section{Introduction}
The crypt of the large intestine is a widely used model for studying
the division of stem cells, for observing the differentiation and,
ultimately, death of cells arising from those stem cells, and for
the genesis of tumors resulting from abnormalities of cell division
and/or death and/or migration. We have previously set up simple
models of cell birth, differentiation and death in the colonic crypt
and used these to analyze parameters of cell behavior which led to
stable equilibria and to tumorigenesis if those parameters were
altered. Our results showed that the classical exponential growth
of an expanding tumor clone could occur, but that
 mutations with weaker effects, particularly on programmed cell death (PCD), could cause tumor growth to a new equilibrium or plateau level of cells. We
speculated that such a form of tumor growth might be particularly
applicable to benign lesions which rarely progress to malignancy,
such as hyperplastic polyps of the colorectum, or lipomas of the
skin.

The model which we originally described was based on a commonly
used simplification of the cell populations within the colonic
crypt (see table 1 of \cite{[TB]}). A population  of stem cells
(number $x_G$ at generation $G$) was assumed to replicate at a
specified rate (division time $t_o$). Each stem cell underwent PCD
with probability $\alpha_1$, renewed itself with probability
$\alpha_3$ , or progressed to a state of semi-differentiation
($p = \alpha_2$). Usually, $x_G$ was assumed to be constant, with
$\alpha_3$ set to 0.5. For the population of semi-differentiated
cells (the size of which  at generation $G$-th will be indicated
as $y_G$ ), division could lead to PCD, renewal or progression
to a fully differentiated state (the size of which at generation
$G$-th will be indicated as  $Z_G$). Fully differentiated cells
underwent PCD with a probability $\gamma$. Mathematically, following the above assumptions,
the dynamics of the cell population was ruled by the linear equations:
\begin{eqnarray}\label{ModLin}
x_{G+1}&=& 2 \alpha_3 x_G \nonumber\\
y_{G+1}&=& 2 \beta_3 \frac{t_o}{t_1} y_G + 2 \alpha_2 x_G \\
z_{G+1}&=& 2 \beta_2 \frac{t_o}{t_1} y_G + (1-\gamma \frac{t_o}{t_2} ) z_G \nonumber
\end{eqnarray} where, of course: $\alpha_1+\alpha_2+\alpha_3=1$ and $\beta_1+\beta_2+\beta_3=1$. If we assume that $x_G$ is constant, it has to be that $\alpha_3=0.5$
and the first equation reads $x_{G+1}=x_G$. When $ 2 \beta_3
\frac{t_o}{t_1} <1 $ the number of semi-differentiated cells
evolves to the following equilibrium value $ y_{eq} = (2 \alpha_2
x_G)/(1-2 \beta_3 \frac{t_o}{t_1}) $, with $y_{eq}$ proportional
to $x_G$. Instead when  $ 2 \beta_3 \frac{t_o}{t_1} \ge 1$ there
is an exponential increase  $y_G \rightarrow +\infty $ as $G
\rightarrow +\infty $. Therefore, $\beta_3$  is a bifurcation
parameter \cite{[Hale0],[Wiggins]}, that is, there exists a
critical value $$ K = \frac{ t_1 }{2 t_o}     $$ such that the
behavior of the system for $\beta_3 < K$  (that is, $y_G
\rightarrow y_{eq}$) is not equivalent  to the behavior for
$\beta_3 \ge K$ ( $y_G \rightarrow +\infty $). Clearly, $\alpha_3$
is also a bifurcation parameter. Note, however, that if $\alpha_3
= 0.5$, the constant number of stem cells and the rate of
progression to a semi-differentiated state are not bifurcation
parameters. If they have a small variations, the behavior of
$y_G$ does not change qualitatively and has only a small numerical
variation.

In this manuscript, we
extend our model to include situations in which there is random fluctuation in cell numbers within each compartment and to situations in which the probabilities of PCD, differentiation or
renewal depend on the number of cells in each compartment.
Incorporating these assumptions introduces non-linearity into the
model, with consequences for the maintenance of stable equilibria
within the crypt cell populations.
\section{Non-linear extension of the model}
\begin{figure}
     \begin{center}
      \leavevmode
        \includegraphics[width=2.4765in,height=1.7355in]{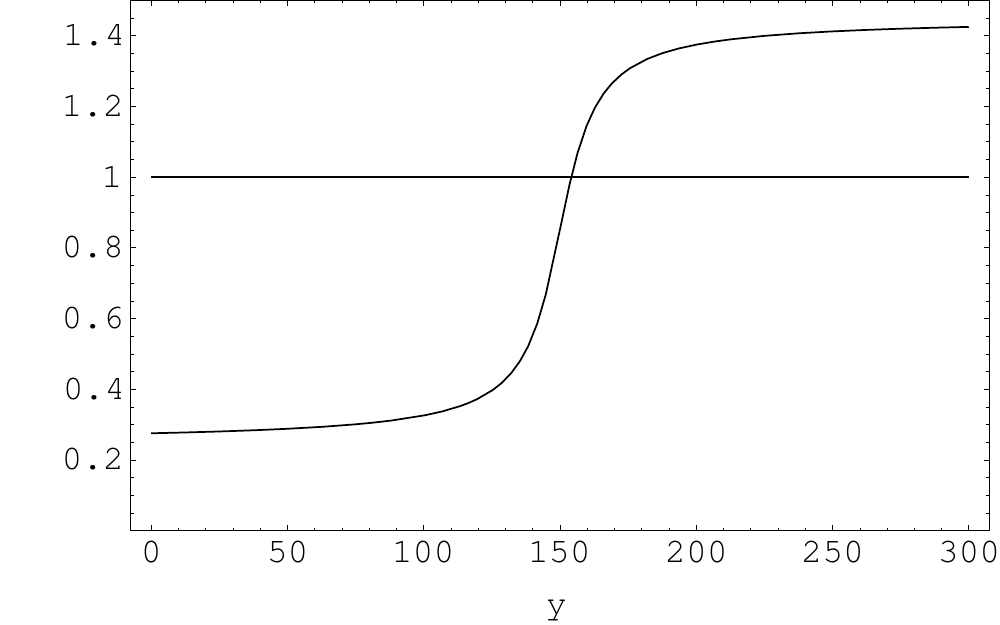}
         \includegraphics[width=2.4765in,height=1.7355in]{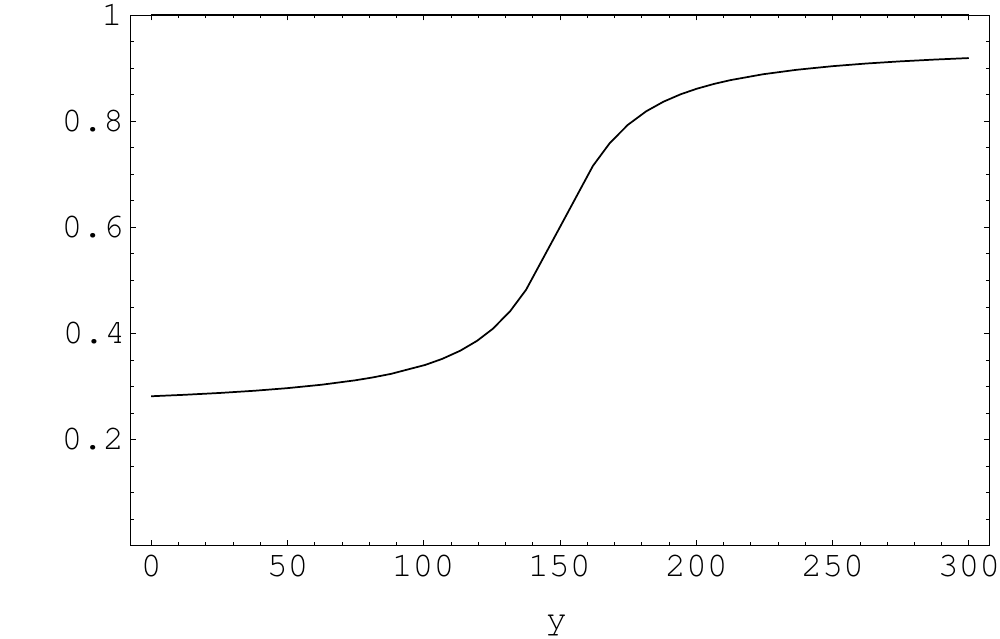}
\caption{Non-linear model of section 4: plots of $2 \frac{t_o}{t_1} \beta_3(y)$. Upper: $ 2 \frac{t_o}{t_1} \beta_3(+\infty) >1$, lower $ 2 \frac{t_o}{t_1} \beta_3(+\infty) <1$}
 \label{betaplots}
     \end{center}
\end{figure}

Let us assume that some function of the number of
semi-differentiated cells may increase or reduce the probability
of PCD. In the former case, the system has, as in the linear
model, only one global asymptotic equilibrium  $y_{eq}$. As a consequence, the
parameter $ 2  (t_o/t_1) \beta_3(y) $ is a  decreasing function of
$y$, and it may be seen that $y_{eq}$ is smaller than the value $ 2 \alpha_2 / X (1- 2
\beta_3 (t_o/t_1) ) $  which would be reached if $\beta_3$ were
constant. This situation may represent normal homeostasis. In
contrast, let us consider the latter case, which represents an
abnormal situation, perhaps one  which results from mutation in
the stem cell compartment. In the latter case, we shall allow
$\beta_1(y)$  to be a decreasing function of  $y$, whereas we will
consider , $\beta_2(y)$ constant or slowly growing, so that $\beta_3(y) $  is a growing function of the
number of semi-differentiated cells. Furthermore we consider that
for low $y$ the variation of that function is slow, such that it
initially is approximately constant. Because of their nature
$\beta_1,\beta_2,\beta_3 \in (0,1)$ and therefore these parameters
shall have asymptotic values, as in figure \ref{betaplots}.
 In practice the $\beta$ functions model a sort of loose threshold mechanism. We do not place a particular constraint on the analytical form of these functions,
since the results which we shall present are qualitatively
dependent only on the "shape" of $\beta_3(y) $. For the
simulations which we shall propose we shall use : $$ \beta_3(y) =
\beta_{3_{min}} + 0.5 \beta_{3_{Max}} \Big( 1 + \frac{2}{\pi}
Atan\big( A(y-y_m)\big) \Big), A>>1$$ or$$\beta_3(y)=\begin{cases}
\beta_{3_{min}} & \text{if $y<y_1$ },\\
\beta_{3_{min}}+\frac{\beta_{3_{Max}}-\beta_{3_{min}}}{y_2-y_1}(y-y_1)
& \text{if $y_1\le y \le y_2$ },\\ \beta_{3_{Max}} & \text{if
$y<y_1$ }.
\end{cases}$$
   Thus instead of the linear set of equations (\ref{ModLin}) we have the following nonlinear discrete dynamical system:
\begin{eqnarray}\label{ModNonl}
x_{G+1}&=& 2 \alpha_3 x_G \nonumber\\
y_{G+1}&=& 2  \frac{t_o}{t_1} \beta_3(y_G) y_G + 2 \alpha_2 x_G \\
z_{G+1}&=& 2  \frac{t_o}{t_1} \beta_2(y_G) y_G + (1-\gamma \frac{t_o}{t_2} ) z_G \nonumber
\end{eqnarray}
Note that the behavior of $z_G$, being $ 0< (1-\gamma
\frac{t_o}{t_2} ) <1$, is determined by the behavior of   $y_G$:
\begin{itemize}
\item When $y_G \rightarrow +\infty $ also $z_G \rightarrow +\infty $;
\item when $y_G \rightarrow + y_{eq} $ also $z$ tends to an equilibrium point.
\end{itemize}
Therefore, in the next section, we shall deal only with the
dynamics of $x_G$ and $y_G$.
\section{ Non-linear model: fluctuation in the number of semi-differentiated cells depends on stem cell number and on $\alpha_2$}
In this section we shall set $$ U = 2 \alpha_2 x_G$$ and consider
it as a static parameter in the input to the equation for the
semi-differentiated cells. The equilibrium equation reads $$ y  =
2 \beta_3(y) (t_o/t_1) y + U$$ which we shall rearrange in the
more convenient form:
\begin{equation}\label{bife} 2 \beta_3(y) (t_o/t_1)= 1 -  \frac{U}{y} \end{equation} that
is, the equilibrium is determined by the intersection of the  $ 2
\beta_3(y) (t_o/t_1)$ curve with the family of hyperbolae $1 -
(U/y)$. The dynamics of the model are essentially determined by
the asymptotic value $ 2 \beta_3(+\infty) (t_o/t_1)$.
\subsection{ Excess self-renewal of semi-differentiated cells}\label{exc}
\begin{figure}
     \begin{center}
      \leavevmode
         \includegraphics[width=2.4765in,height=1.7355in]{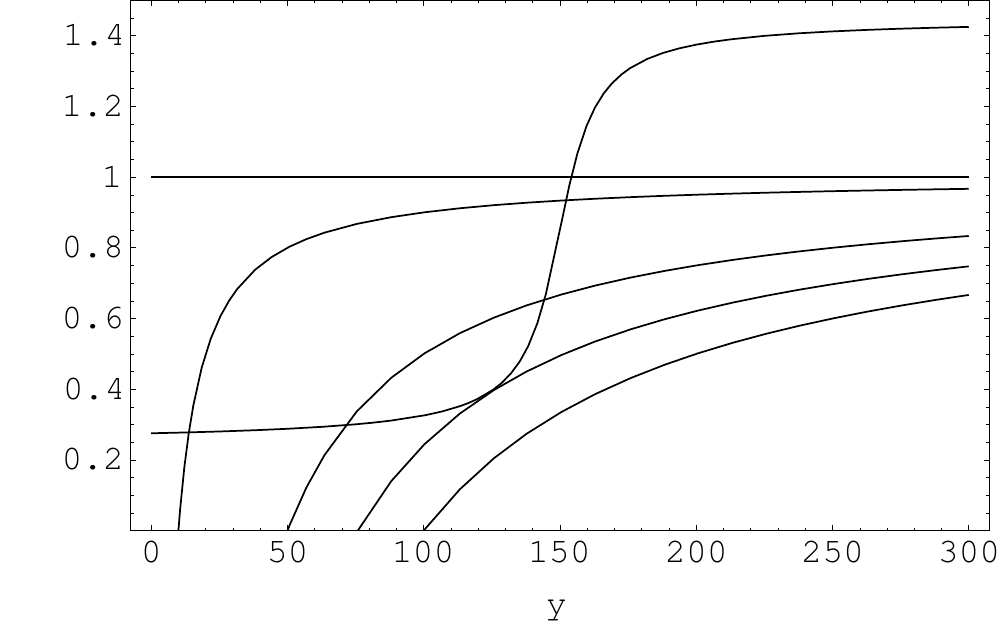}
       \caption{ Excess of self-renewal of semi-differentiated cells ($ 2
\frac{t_o}{t_1} \beta_3(+\infty) >1$). There are 2 (for $U<U_m$) or
or 0 ($U>U_m$) equilibria. In the limit case $U=U_m$ there is
tangency (a double equilibrium point). The higher the values of
$U$, the further to the right are the hyperbolae.}
      \label{SaddleNodeBeta}
     \end{center}
\end{figure}

\begin{figure}
\centering
\subfigure[ ]{\includegraphics[width=2.4765in,height=1.7355in]{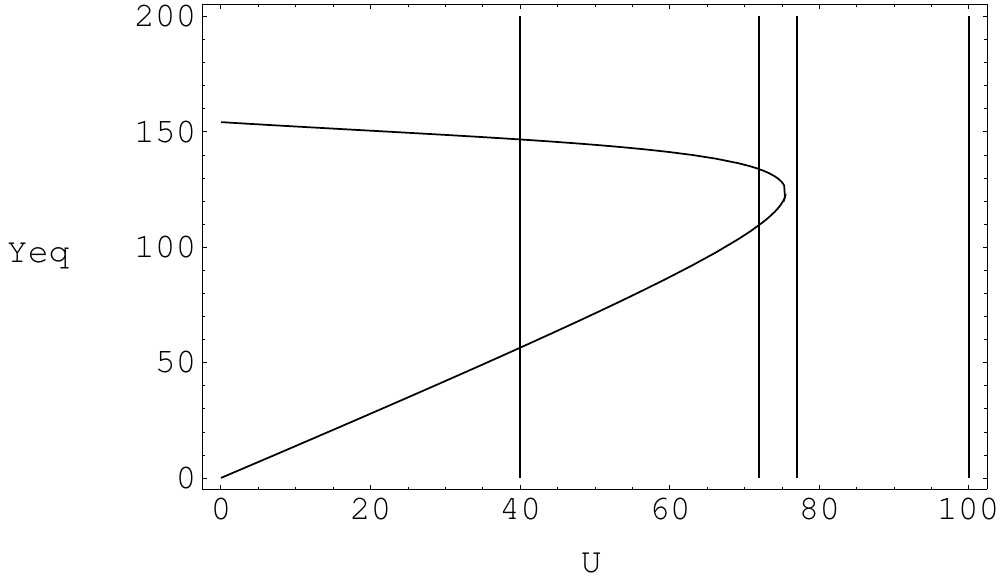}}\\
\subfigure[]{\includegraphics[width=2.4765in,height=1.7355in]{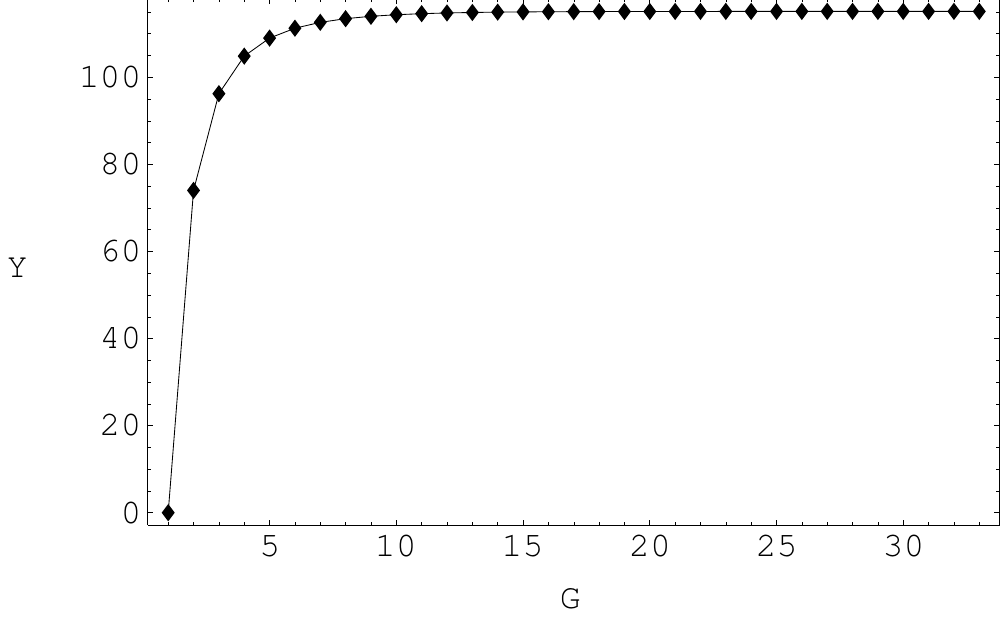}}\quad
\subfigure[ ]{\includegraphics[width=2.4765in,height=1.7355in]{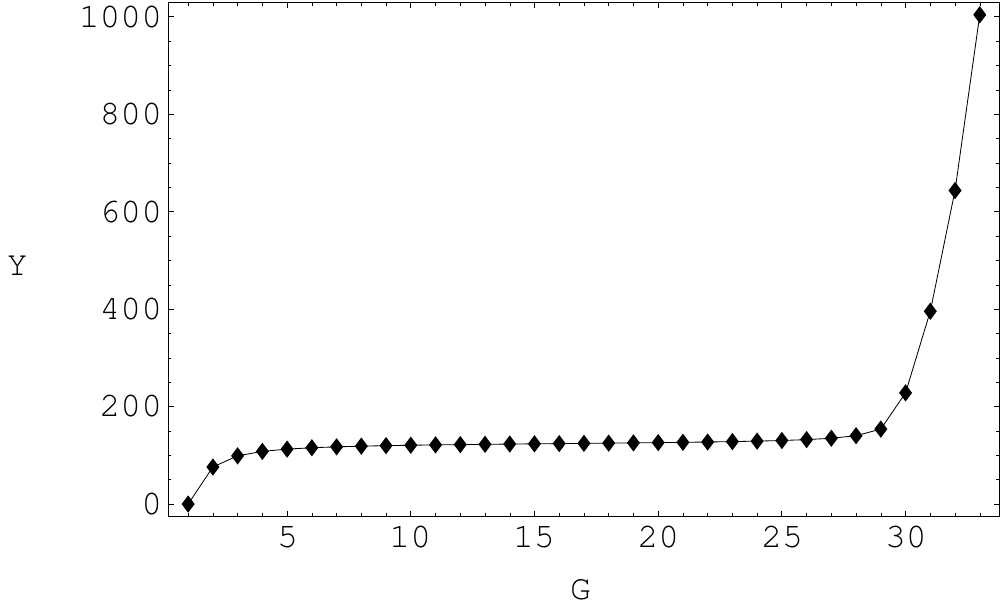}}%
 \caption{(a): Typical saddle node diagram corresponding to
figure \ref{SaddleNodeBeta}: for $U<U_m$ there are 2 equilibria
(one stable, the lower, and one unstable)  and for $U>U_m$ no
equilibria exist, (b): simulation corresponding to a value
of $U$ slightly less than the critical value $U_m$; (c):
simulation corresponding to a value of $U$ slightly greater than
the critical value $U_m$, which implies exponential
explosion.}\label{SaddleNodeAndPlots}
\end{figure}

Here $$  2 \beta_3(+\infty) (t_o/t_1) > 1.$$ As a consequence, there
exists (see Figure \ref{SaddleNodeBeta}) a threshold value $U_m$ for
$U$, such that for each $U \in (0,U_m)$ there are two points of
intersection, (that is, two equilibria); for $U=U_m$, the lines
collide, and for $U>U_m$ there are no equilibrium points. In the
limit case $U=U_m$ there is tangency (a double equilibrium point).
The higher the value of $U$, the more to the right are the
hyperbolae. From a static point of view (that is, considering
different but constant values of $U$), there is a saddle node
bifurcation \cite{[Hale0],[Wiggins]}. From figure
\ref{SaddleNodeAndPlots}-(a) , we see that for $U<U_m$, there is a
stable lower equilibrium $y^{Stab}(U)$, followed by an unstable
equilibrium $y^{Unst}(U)$; the equilibria tend to collide for $U
\rightarrow U_m$. The point $y^{Stab}(U)$ has an attraction basin
equal to the whole range $\big(0, y^{Unst}(U)\big)$, which is easily
verified by using the following Liapunov-La Salle function $L(y)= |
y -  y^{Stab}(U)|$. For all $ y \in (y^{Unst}(U),+\infty)$, there is
instability and unbounded growth of the number of
semi-differentiated cells ($y \rightarrow +\infty $). From figure
\ref{SaddleNodeAndPlots}, the critical threshold value corresponds
to the maximum of the curve $U=U(Y_{eq})$ $=$ $ Y_{eq}(1-2 (t0/t1)
\beta_3 (Y_{eq}) ) $. As a consequence $U_m =Y_m(1-2 (t0/t1) \beta_3
(Y_{m}) )$ where $Y_m$ is the unique real solution of the null
derivative equation $ \frac{dU(Y)}{dY}=0 \Rightarrow 2 \frac{t0}{t1}
(\beta_3 (Y)+ \frac{d \beta_3 (Y)}{dY}) = 0 $ From figures
\ref{SaddleNodeBeta} and \ref{SaddleNodeAndPlots}-(a), it is evident
that the first, stable equilibrium point is proportional to $U$ for
small and also "moderately small" values of this parameter, since
the first intersection between the hyperbola $1-U/y $ and the curve
$2 (t0/t1) \beta_3 (Y) $ is located in the a zone in which the
second curve is approximately constant. Let us simulate what happens
in the proximity of the bifurcation value $U_m$. Figure
\ref{SaddleNodeAndPlots}-(b) shows a simulation of a system having $
U $ slightly lower than $ U_m $. It shows classical behavior,
initial growth followed by a plateau. There is no qualitative
difference between this behavior and that shown in figure 2 of
\cite{[TB]}. A more interesting situation is shown in figure
\ref{SaddleNodeAndPlots}-(c), in which $U$ is slightly greater than
$U_m$ and in which the plateau is only temporary, since it is
followed by an exponential "explosion". A small positive variation
around $U_m$, therefore, makes the system grow without bound.
\subsection{Random fluctuation in cell numbers}\label{Random fluctuation in cell numbers} We have analyzed the bifurcation of the growth in cell
numbers from a static point of view, for given values of the
parameter $U$, that is, for given values of the number of stem
cells and of the rate of progression from stem cell to
semi-differentiated cell. In reality, the bifurcation may be
driven by dynamic phenomena, such as variation  - even if
transient - in the number of stem cells. The rate of PCD or the
rate of progression to semi-differentiated cells may also vary. We
shall simulate such variability by means of the following
non-linear, stochastic, discrete dynamic system, using the
original model
\begin{eqnarray}\label{ModNonlStoch}
x_{G+1}&=& 2 \alpha_3(G) x_G \nonumber\\
y_{G+1}&=& 2  \frac{t_o}{t_1} \beta_3(y_G) y_G + 2 \alpha_2(G) x_G \\
z_{G+1}&=& 2  \frac{t_o}{t_1} \beta_2(y_G) y_G + (1-\gamma \frac{t_o}{t_2} ) z_G \nonumber
\end{eqnarray}
except that $\alpha_3(G)$ and $\alpha_2(G)$  are now stochastic
processes. First, we simulated a sporadic random increase in the
number of stem cells near the bifurcation by considering a
constant $\alpha_2$  such that $$\alpha_2 x_1 = U_m (1-\omega),
\omega << 1$$ that is, $U$ is slightly under the bifurcation
value. We assume a stochastically varying $\alpha_3$, which will
follow this rule: $$\alpha_3(G)=\begin{cases} 0.5& \text{if $G<G_1$
or $G> G_2$ },\\ 0.5+\eta & \text{if $G \in [G_1,G_2]$ }.
\end{cases}$$with $G_1$,$G_2$ and $\eta$ random variables and such that
$\alpha_2 x(G_2) > U_m$  . The result patterns are identical
(exponential growth) in each simulation, but with different
start-points for the exponential 'explosion' in growth. Therefore,
this model seems to indicate that, since the exponential growth
may be due to saddle-node bifurcations (from a steady state to an
unbounded growth) and since these phenomena may be driven by a
small noise, the initial time in which this growth (for example,
that of a tumor) starts is not calculable in a deterministic way.
We also undertook different stochastic simulations, in which we
assumed that $\alpha_2(G) $ , $\alpha_3(G) $  are gaussian or
uniform random variables with $E[\alpha_3(G)] = 0.5 $. The
standard deviation was variable and such that for a finite number
of generations $Prob( U_G > U_m ) $ is significantly greater than
$0$.  The mean values of parameters are such that, if the system
were deterministic, it would have a plateau (that is, $E[ U ] <
U_m$). The simulations showed that all outcomes do not result in
exponential 'explosion' in a given finite temporal window.
Intuitively, this means that the random effects do not
sufficiently 'scramble' the system to drive it into the zone of
instability. More formally, given the stochastic process $U_G=
\alpha_2(G)x_G$, there is exponential growth only if, at the
generation $G$ when $U(G)$ assumes its minimum value $\hat{U}$,
the number $y(G)$ of semi-differentiated cells is such that $y
\in (y^{Unst},+\infty) $. We performed stochastic simulations for
various mean values of $U$ such that $E[ U ] < U_m$, constant
$\alpha_2$ and each $\alpha_3(G)$ gaussian with mean 0.5 and
$\sigma = 0.05$ ( which implies that $E[U]=2 \alpha_2 X_0$ ). The
simulations showed that also for low values of $E[U]$  there may
be exponential 'explosion' in a  fixed number of generations (for
ex. 100) and that  for $G \rightarrow +\infty$ there is
exponential 'explosion' in all cases.

The possible behaviors with random variations of $U$ can be
summarized as follows:
\begin{itemize}
\item   if $U=U_1< U_m$  increases in a random generation $G$ to an upper value $U_2 > U_m$, the system loses its equilibrium and exponential growth starts;
\item   if $U$ varies randomly in a range $[a,b]$, with $b<U_m$, there is no loss of equilibrium;
\item   if $U$ varies randomly for a finite number of generations in a range $[a,b]$, with $b>U_m$ , there may or may not be loss of equilibrium;
\item if $U$ varies randomly and  permanently in a range $[U_o-c,U_o+c]$, with $c>0$ and $ E[U]=U_o \ge U_m$ , there is loss of equilibrium in the long
run;
\item if $U$ varies randomly and  permanently in a range $[U_o-c,U_o+c]$, with $c>0$ and $U_o < U_m$ , we may find an $\epsilon > 0$  such that
if $c>(U_m - U_o)(1+\epsilon)$ there is loss of equilibrium in the
long run, whereas if $c < (U_m - U_o)(1+\epsilon)$, $y_{+\infty}$ remains bounded.
\end{itemize}
\subsection{No excess renewal of semi-differentiated
cells}\label{noexc}
\begin{figure}
     \begin{center}
      \leavevmode
         \includegraphics[width=2.4765in,height=1.7355in]{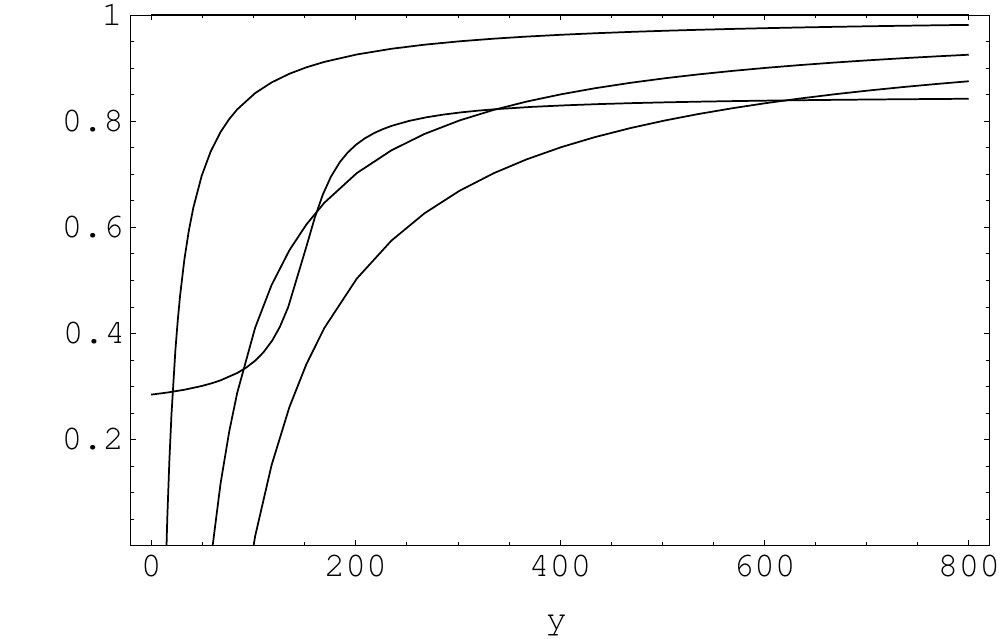}
       \caption{No excess of self-renewal of semi-differentiated cells
($ 2 \frac{t_o}{t_1} \beta_3(+\infty) < 1$). There are 1 (for
$U<U_m$ and for $U>U_M$) or 3 ($U_m < U<U_M$) equilibria. In the
limit cases $U=U_m$ and $U=U_M$  there is tangency (a double
equilibrium point). The higher the values of $U$, the further to
the right are the hyperbolae.}
\end{center}\label{HystBeta}
\end{figure}

\begin{figure}
\centering
\subfigure[ ]{\includegraphics[width=2.4765in,height=1.7355in]{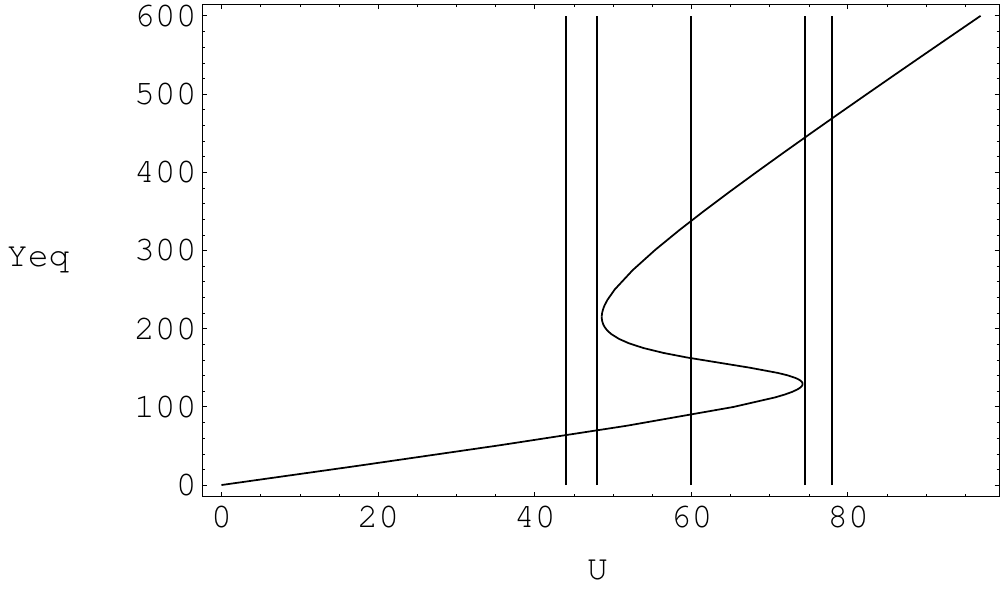}}\\
\subfigure[ ]{\includegraphics[width=2.4765in,height=1.7355in]{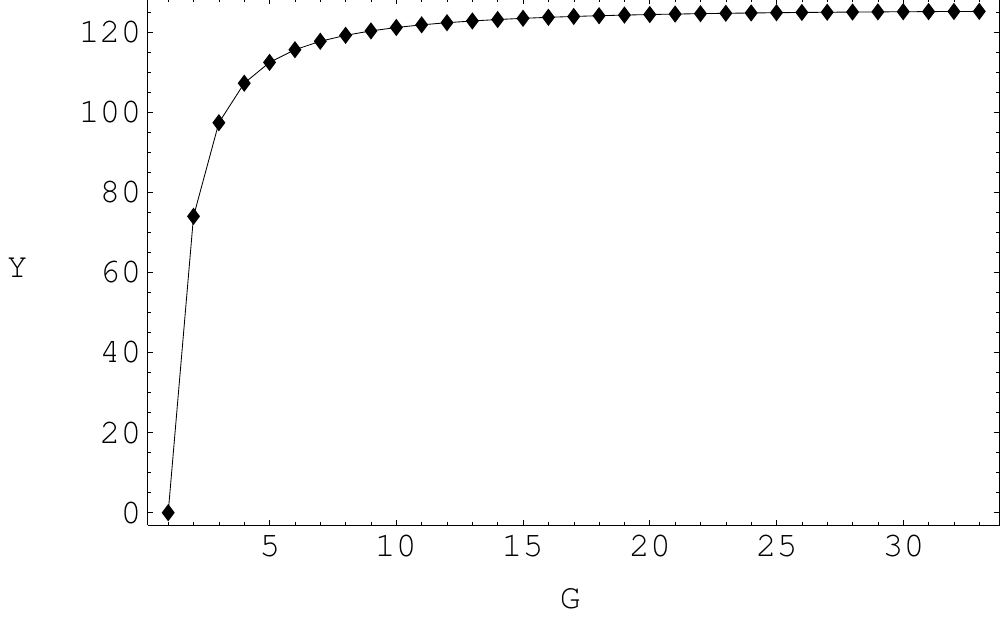}}\quad
\subfigure[ ]{\includegraphics[width=2.4765in,height=1.7355in]{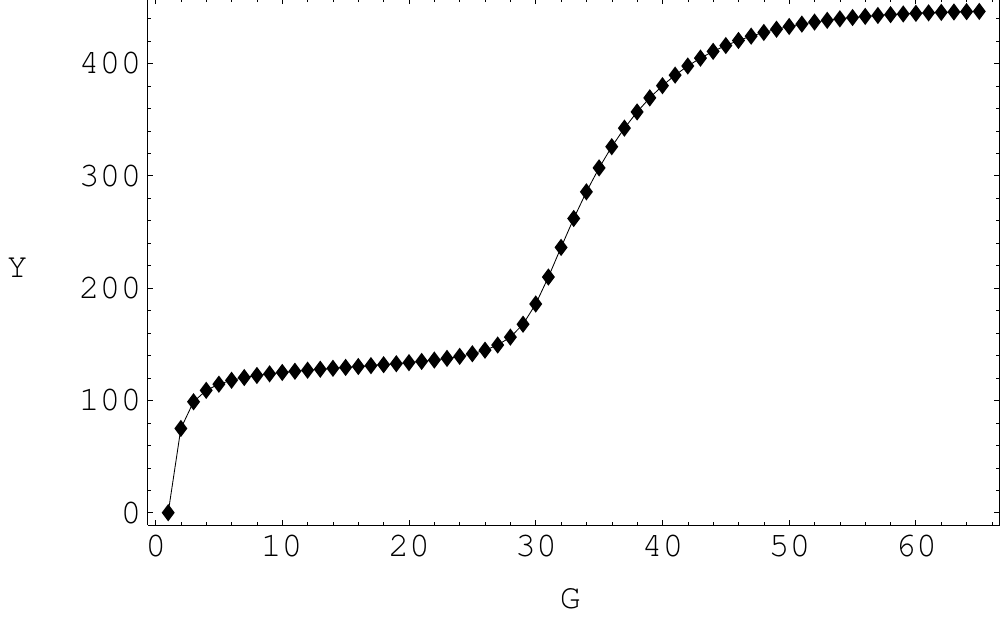}}%
 \caption{(a): Typical Hysteretic Global Bifurcation diagram: 2
local saddle node in a S-shaped curve (associated with fig.
\ref{HystBeta}). Simulations corresponding to (a): a
value of $U$ slightly less than the critical value $U_M$; (b) a
value of $U$ slightly greater than the critical value $U_M$.}
\label{hysteresis}
\end{figure}

Here: $$  2 \beta_3(+\infty) (t_o/t_1) < 1$$ As a consequence, as
shown in figure \ref{HystBeta}, for increasing values of $U$ the
system passes from $EQ1$ to $EQ3$ and back to $EQ1$  equilibrium
points. This is a typical situation which leads to a hysteresis
bifurcation diagram [4], as in figure \ref{hysteresis}-(a). It is
possible to determine $2$ particular values for $U$, let us call
them $U_{\mu}$  and $U_m$ with $U_m > U_{\mu} $, such that:
\begin{itemize}
\item For $U<U_{\mu}$ there is one globally asymptotically stable equilibrium,  $y_{low}(U)$;
\item For $U>U_{m}$ there is one globally asymptotically stable equilibrium,  $y_{high}(U)$;
\item For $ U_{\mu}<U<U_{m}$  there are three coexisting equilibria: a new third unstable equilibrium $y_{mid}(U)$ (with: $y_{low}(U) < y_{mid}(U) < y_{high}(U) $  ); and the two previous equilibria: $y_{low}(U)$ (with basin of attraction now: $[0, y_{mid}(U) )$ ) and $ y_{high}(U) $ (with basin of attraction: $( y_{mid}(U) ,+\infty)$).
\end{itemize}
The above stated stability properties are easily proved by using these two Liapounov-La Salle functions: $V_{low}(y)= | y- y_{low}(U)|$ and
$V_{high}(y)= | y- y_{high}(U)|$.

With reference to the bifurcation diagram, $U_{\mu}$ and $U_m$
correspond, respectively, to the minimum and maximum of the
bifurcation curve $U=U(Y_{eq})$. In this case,  if we call $Y_{\mu}$
and $Y_M$ (with $Y_M < Y_{\mu}$  )  the two solutions of the
equation $dU/dY = 0$, we obtain $$ U_{\mu} = Y_{\mu} (1-2 (t0/t1)
\beta_3(Y_{\mu})), U_m= Y_M (1-2 (t0/t1) \beta_3(Y_M)$$ By using the
terminology of catastrophe theory, when $U$ varies and become the
greater (lower) than the value of $U_m$ ($U_{\mu}$), there is an
elementary catastrophe, (that is, with an infinitesimal variation of
a control parameter, $U$ in our case, there is a finite variation in
the equilibrium point). In figure \ref{hysteresis}, the effect of
the hysteresis bifurcation on cell numbers is shown by means of two
simulations: in particular in figure \ref{hysteresis}-(b),
corresponding to a value of $U$ slightly greater than $U_m$, the
system appears to have reached its equilibrium, but restarts,
increasing to reach a new and higher asymptotic value.

\begin{figure}
\centering
\subfigure[ ]
{
    \label{fig:sub:a}
    \includegraphics[width=2.4765in,height=1.7355in]{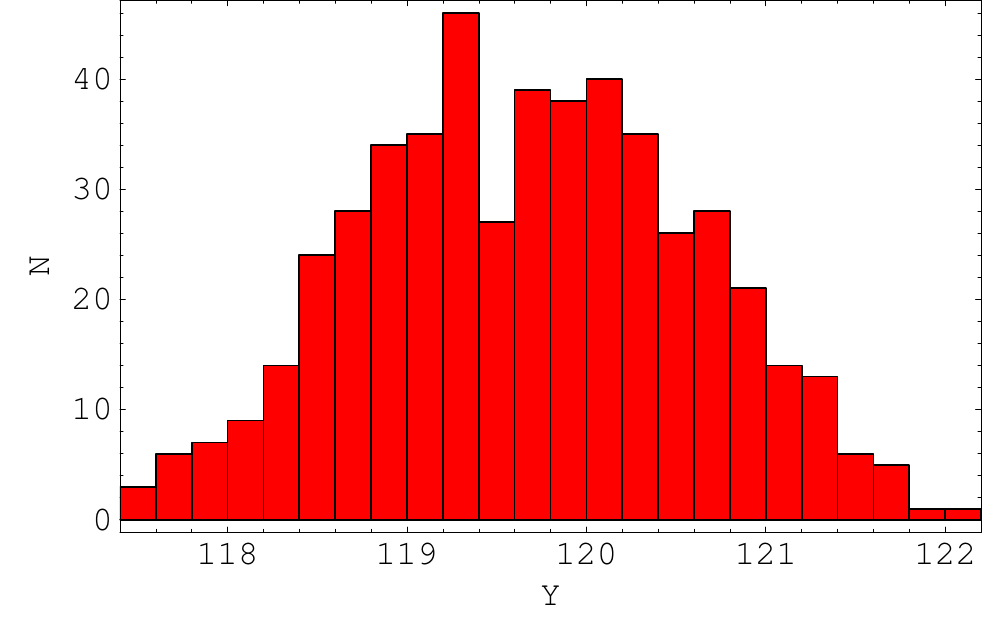}
}
\hspace{1cm}
\subfigure[ ]
{
    \label{fig:sub:b}
    \includegraphics[width=2.4765in,height=1.7355in]{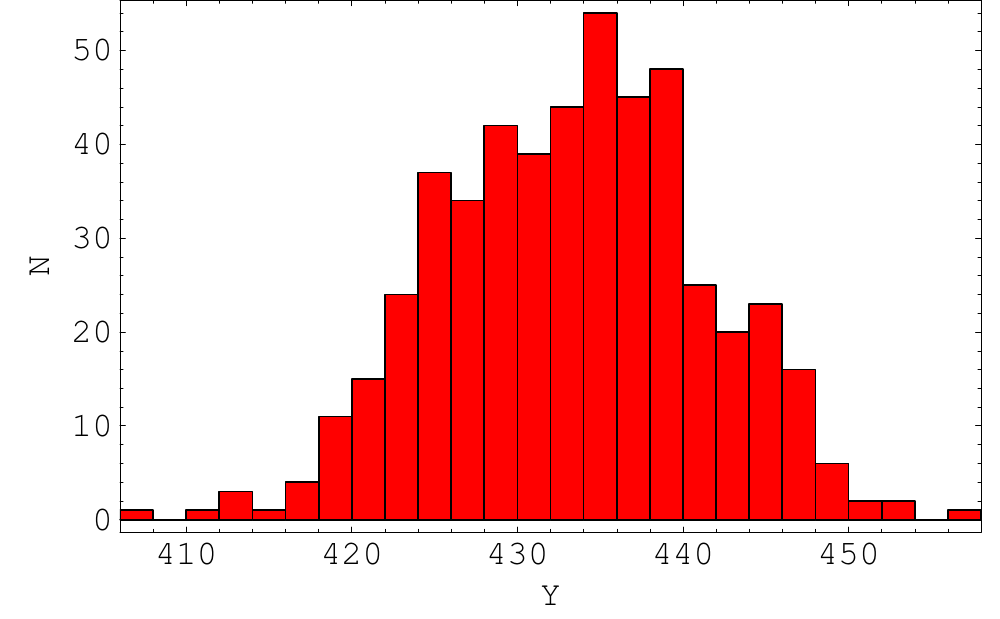}
}
\caption{Histogram probability density for $Y_{1000}$, relative to the hysteresis bifurcation diagram
of figure \ref{hysteresis} with $U_o=73$ and (a): $c=1$, (b):
$c=7$. Note that the two densities are located in different ranges
of values of $y$.}\label{histograms}
\end{figure}

Furthermore, in the case in which the deterministic model has two, co-existing, alternative,
stable equilibria, the effect of stochastic variation of
parameters is a catastrophic noise-induced transition
\cite{[Horst]} from a lower (upper) to an upper (lower)
equilibrium point (figure \ref{hysteresis}). Let us suppose
constant $\alpha_2$  and stochastically varying $\alpha_3$ in a
way such that $U_G \in [U_o-c,U_o+c]$,$c>0$. As $G \rightarrow
+\infty$, the system will have the following behaviors:
\begin{itemize}
\item   if $c<U_m - U_o $, then the asymptotic value $y_{+\infty}$ will be a random variable with a probability
density $ \rho(y; c) $ (where we stress the dependence of the density on the parameter $c$) having a maximum corresponding to a low equilibrium point as in figure \ref{histograms}-(a);
\item if $U_o-c<U_m$ and $U_o>U_m$ we have a $\rho(y; c)$ such that its maximum is near to an upper equilibrium point.
\item when $c>U_m - U_o $, we may find an $\epsilon>0$ such that if $c<(U_m - U_o)(1+\epsilon)$ the above density remains unchanged, whereas if $c>(U_m - U_o)(1+\epsilon)$ the above probability density $\rho(y; c)$ changes, since its maximum is now near to an upper equilibrium point, as in figure \ref{histograms}-(b).
\end{itemize}
This changing in the probability density distribution of
$y_{\infty}$ (as well as the change in the behaviors summarized at
the end of section \ref{Random fluctuation in cell numbers}) has
been called noise-induced transition \cite{[Horst]} or, in recent
literature, stochastic bifurcation \cite{[Arnold]}. From a
computational point of view, in practice we approximate
$\rho(y_{\infty}; c)$ with the probability density distribution of a
$y_G$ with $G>>1$. It results that the nearer $c $ is to $U_m - U_o
$, the greater the value of $G$  which has to be chosen. On the
contrary if one establishes a priori a finite number of generations
$\hat{G}$, there are two values $c_1>U_m - U_o$ and $c_2>c_1$ such
that if $c<c_1$ the probability density of $y(\hat{G})$ has a low
peak, if $c>c_2$ it has an high peak and if $c_1 < c < c_2$ it has
two peaks: one low and one high.
\subsection{Mixed behaviors}\label{mix}
Remaining in the framework of increasing $\beta(y)$, there may be
the theoretical possibility to have more complex configurations. In
fact, if $\beta(y)$ has $m \ge 2$ inflexion points and if it is $2
(t_o/t_1) \beta(+\infty) > 1$, there may be $n \le m$ stable
equilibria and $n$ unstable equilibria (in alternated sequence
starting from a stable equilibrium: STABLE , UNSTABLE, STABLE,
UNSTABLE,etc$\dots$). Proceeding as in the sections \ref{exc} and
\ref{noexc}, it is easy to show that there are $n-1$ hysteresis
bifurcations followed by a final saddle-node bifurcation. Thus also
in this more complex case, it is possible to find a threshold value
$U_m$ exceeding which there is exponential explosion. When $2
(t_o/t_1) \beta(+\infty) < 1$, there may be $n+1$ stable equilibria
and $n$ unstable equilibria, and $n$ hysteresis bifurcations.
Summarizing, also in these more complex case, the biological
findings previously illustrated do not change, since the behavior is
simply a mix of the behaviors illustrated in the two previous
subsections.
\section{Non-linear model: contribution to the rate of PCD by both the semi-differentiated and the differentiated cells}
The first natural extension to the proposed nonlinear model is to
consider that both semi-differentiated and fully differentiated
cells may contribute to the variability in PCD. Therefore we have
$$ \beta_1(y,z),\beta_2(y,z),\beta_3(y,z)$$ with: $$
\frac{\partial \beta_1}{\partial y } < 0, \frac{\partial
\beta_1}{\partial z } < 0, \frac{\partial \beta_2}{\partial y }
\ge 0, \frac{\partial \beta_2}{\partial z } \ge 0, \frac{\partial
\beta_3}{\partial y } >0, \frac{\partial \beta_3}{\partial z }> 0
$$ For example, by assuming that there is no difference in the
contribution of the semi-differentiated and fully differentiated
cells, a natural candidate function is $$ \beta_3^o(y+z) $$ where $
\beta_3^o(.) $   is a function shaped as the one in figure
\ref{betaplots} (different contribution may be modeled by assigning
to $y$ and $z$ two different positive weights: $ \beta_3^o(w_y y+
w_z z)$, e.g. $w_y = 0$ and $w_z = 1$ i.e. dependence only on the
fully differentiated cells). The following model results
\begin{eqnarray}\label{ModNonlCoop}
x_{G+1}&=& 2 \alpha_3 x_G \nonumber\\ y_{G+1}&=& 2 \frac{t_o}{t_1}
\beta_3(y_G + z_G) y_G + 2 \alpha_2 x_G \\ z_{G+1}&=& 2
\frac{t_o}{t_1} \beta_2(y_G + z_G) y_G + (1-\gamma \frac{t_o}{t_2}
) z_G \nonumber
\end{eqnarray}
Note that: $$ \frac{\partial y_{G+1}}{\partial z_G } >0 ,
\frac{\partial z_{G+1}}{\partial y_G } \ge 0 $$ that is,
mathematically, the system may be classified as cooperative, in
agreement with what we suggested biologically . Defining the new
variable: $$ s_G = y_G + z_G $$ (note that $s_G$ it is such that
$s_G \ge y_G$ and $s_G \ge z_G$) and summing the third and the
second equation of (\ref{ModNonlCoop}) we obtain (for constant
$U=\alpha_2 x_G$):
\begin{eqnarray}\label{ModNonlCoopSum}
y_{G+1}&=& 2  \frac{t_o}{t_1} \beta_3(s_G) y_G + U \\
s_{G+1}&=& 2  \frac{t_o}{t_1} ( \beta_2(s_G)+\beta_3(s_G)) y_G + (1-\gamma \frac{t_o}{t_2} ) (s_G-y_G)+U \nonumber
\end{eqnarray}
whose equilibrium points are determined by the equation
\begin{equation}\label{bifcoop}
2\frac{t_o}{t_1}\beta_3(s) = 1 - \frac{U}{s}
\big(1+2*\frac{t_2}{\gamma t_1} \beta_3(s) \big)
\end{equation}
which is very similar to equation (\ref{bife}) and which leads to
the same bifurcation curves. Simulations did not show a
qualitative difference between non-cooperation and cooperation.
From a quantitative point of view, on the contrary, there are
differences: the dynamics are faster in the co-operative model, in
agreement with the fact that $\beta_3$ depends on the sum $y+z$.
\section{Dependence of the parameters on ratio of semi-differentiated to stem cells}
If we suppose that the parameters do not depend directly on $y$,
but that they depend, for example, on the ratio: $$\frac{y}{x} $$
we obtain, because of the constance of the $x_G$, similar results
as far as concerns the bifurcation diagrams. In fact, if we
define the new variable:
\begin{equation}\label{ratio1} \varrho = \frac{y}{X}\end{equation}
we may write the equilibrium equation in the form:
\begin{equation}\label{bifratio}
2\frac{t_o}{t_1}\beta_3(\varrho) = 1 - \frac{2 \alpha_2}{\varrho}
\end{equation}
the only difference in this case being that the bifurcation
parameter is no longer $U=2 X \alpha_2$, but is now the parameter
$\alpha_2$ alone. As consequence of (\ref{ratio1}), the
equilibrium value is proportional to $X$. again the same
bifurcation diagram holds also by proposing a dependence on the
following ratio: $$ \frac{y}{y+X} $$ the equilibrium equation may
be written as:
\begin{equation}\label{bifratio2}
2\frac{t_o}{t_1}\beta_3(\frac{\varrho}{\varrho + 1}) = 1 - \frac{2
\alpha_2}{\varrho}
\end{equation}
which leads to the same bifurcation diagram because of the
geometrical properties of the function $\varrho/(\varrho + 1)$ and
$\beta_3$.
\section{Conclusions}
We have presented a development of our model of cell birth,
differentiation and death in the colonic crypt, which was used as a
model for tumorigenesis. Our previous model, which was linear,
showed that failure of PCD - for example, in tumorigenesis - could
lead either to exponential growth in cell numbers or to growth to
some plateau. We note here that, remaining in the linear framework,
some biological noteworthy modifications are possible. In
particular: periodic or random changes of the parameters, in order
to take into the account interactions with the microenvironment;
and, as suggested by one of the referees, incorporating exponential
decay in the growth parameter, as in the Gompertz model
\cite{[Molski]}. This second point, in particular, would deserve a
further analysis, also in the non-linear case.\\ Our new models have
incorporated realistic fluctuation in the parameters of the model
and have introduced non-linearity by assuming dependence of
parameters on the numbers of cells in each state of differentiation,
perhaps as a result of a mutation occurring in and spreading through
the stem cell population. Furthermore, recently in \cite{[Molski]}
it has been proposed for cancer cells a mechanism of self
organization through cell to cell communication similar to the
quorum sensing of bacteria, and other kind of cell-density
dependence of apoptosis has been proposed in \cite{[YC]} and
\cite{[Hardy]}. Our new models are characterized by bifurcation
between increase in cell numbers to stable equilibrium or explosive
exponential growth, although, in particular cases, two coexisting
stable equilibria exist together with an unstable equilibrium (see
sect. \ref{exc}). Moreover, we note here that for more complex (but
increasing) shapes of $\beta(y)$, there may be even other coexisting
equilibria. However, it is easy to show that these more complex
(and quite theoretical) configurations of equilibria do not alter the biological findings of the present work.\\
If we assume fluctuation in cell numbers undergoing PCD (whether
determined or random), the incorporation of non-linearity into the
model generally makes exponential growth of a tumor more likely, as
long as (for random fluctuation), the number (or the proportion) of
cells progressing from a stem cell to a semi-differentiated state
exceeds a certain critical value for a sufficiently long period of
time. In most  - but not all - cases, once exponential growth has
started, this process is irreversible. In some circumstances,
exponential growth may be preceded by a long plateau phase,
mimicking equilibrium, of variable duration: thus apparently
self-limiting lesions may not be so in practice and the duration of
growth of a tumor may be impossible to predict on the basis of its
size.  Our results show that the consequences of failure of PCD are
complex and difficult to predict. Progression of a tumor is not
necessarily caused by acquiring additional extra genetic or
epigenetic changes, but may simply be a consequence of 'normal'
alterations in cell turnover or fluctuations in numbers, owing to,
for example, tissue damage. Apparent regression of tumors may occur
similarly. While failure of PCD is more likely than simple increased
cell replication to be associated with benign, generally
non-progressing tumors, the exponential growth observed suggests
that PCD failure is potentially a general mechanism of
carcinogenesis.
\section{Acknowledgements}
We wish to thank an anonymous referee for her/his valuable comments.

\pagebreak

\pagebreak
\pagebreak

\pagebreak

\pagebreak

\pagebreak

\pagebreak

\end{document}